\documentclass[journal]{IEEEtran}

\usepackage[justification=centering]{caption}
\usepackage{float}
\usepackage{graphicx}
\usepackage{mathtools}
\usepackage{subcaption}
\usepackage{epstopdf}
\usepackage[english]{babel}
\usepackage[utf8x]{inputenc}
\usepackage{amsmath}
\usepackage{tikz}
\usepackage{xspace}
\usepackage{siunitx}
\usepackage{lettrine}
\usepackage{enumitem}
\usepackage{amsmath}
\usepackage{amsfonts}
\usepackage{amssymb}
\usepackage{breqn}
\usepackage{lipsum}
\usepackage{etoolbox}
\usepackage{multirow}
\usepackage{tabularx}
\usepackage{longtable}
\usepackage{siunitx}
\usepackage{dblfloatfix}
\usepackage{float}
\usepackage{xcolor}
\usepackage[font=footnotesize]{caption}

\DeclareCaptionLabelSeparator{periodspace}{.\quad}
\usepackage{cite}
\usepackage{multirow}
\usepackage{hyperref}
\usepackage{graphicx}
\usepackage{amsmath}
\usepackage{multicol}
\usepackage{mathtools}
\usepackage{tabularx}
\usepackage{longtable}
\usepackage{mdwlist}
\usepackage[linesnumbered,ruled,vlined]{algorithm2e}
\usepackage{balance}
\usepackage[square,numbers]{natbib}
\usepackage{booktabs}
\usepackage{algorithmic}

\ifCLASSINFOpdf
  
\else
 
\fi

\begin{document}

\captionsetup[figure]{labelsep=colon,name={Fig.}}

\title{Fault Location Using the Natural Frequency of Oscillation of Current Discharge in MTdc Networks}

\author{Bhaskar Mitra,~\IEEEmembership{Member,~IEEE,}
        Suman Debnath,~\IEEEmembership{Member,~IEEE,}
        and~Badrul Chowdhury,~\IEEEmembership{Senior~Member,~IEEE}
\thanks{Bhaskar Mitra is currently with Idaho National Laboratory, Idaho Falls, ID, 83401, USA, e-mail: bhaskar.mitra@inl.gov.}
\thanks{Suman Debnath is currently with Oak Ridge National Laboratory, Knoxville, TN, 37932, USA, email: debnaths@ornl.gov.}
\thanks{Badrul Chowdhury is currently with the Department of Electrical and Computer Engineering, University of North Carolina, Charlotte, NC, 28223, USA, e-mail: b.chowdhury@uncc.edu.}
\thanks}

\markboth{Journal of \LaTeX\ Class Files,~Vol.~14, No.~8, August~2015}
{Shell \MakeLowercase{\textit{et al.}}: Bare Demo of IEEEtran.cls for IEEE Journals}

\maketitle

\begin{abstract}
This paper discusses a novel fault location approach using single ended measurements. The natural dissipation of the circuit parameters are considered for fault location. A relationship between the damped natural frequency of oscillation of the transmission line current and fault location is established. The hybrid dc circuit breaker (dcCB) interrupts the fault current, whereby the transmission line current attenuates under the absence of any driving voltage source. The line capacitance discharges into the fault at a specific frequency of oscillation and rate of attenuation. Utilizing this information, the fault location in a multi-terminal direct current (MTdc) network can be predicted. A  three  terminal  radial  model  of  a MTdc is used for performance evaluation of the proposed method using Power System Computer Aided Design (PSCAD)/Electromagnetic Transients including dc (EMTdc).
\end{abstract}


\IEEEpeerreviewmaketitle

\section{Introduction} \label{Introduction}

\IEEEPARstart{T}{he} electric grid is undergoing a technological transformation as a result of increasing environmental awareness to reduce carbon emissions. As a result of the push for more renewable energy integration with resources like solar, wind, tidal energy etc. where the generation source is located at a distance far away from load centers, the High Voltage direct current (HVdc) transmission technology has taken prominence over High Voltage alternating current (HVac) transmission \cite{Mitra2018}. Larger power transfer capability, lower power losses and flexible control have made HVdc a popular choice \cite{Mitra2016}.\par

\begin{figure}
    \centering
    \includegraphics[width=0.55\textwidth]{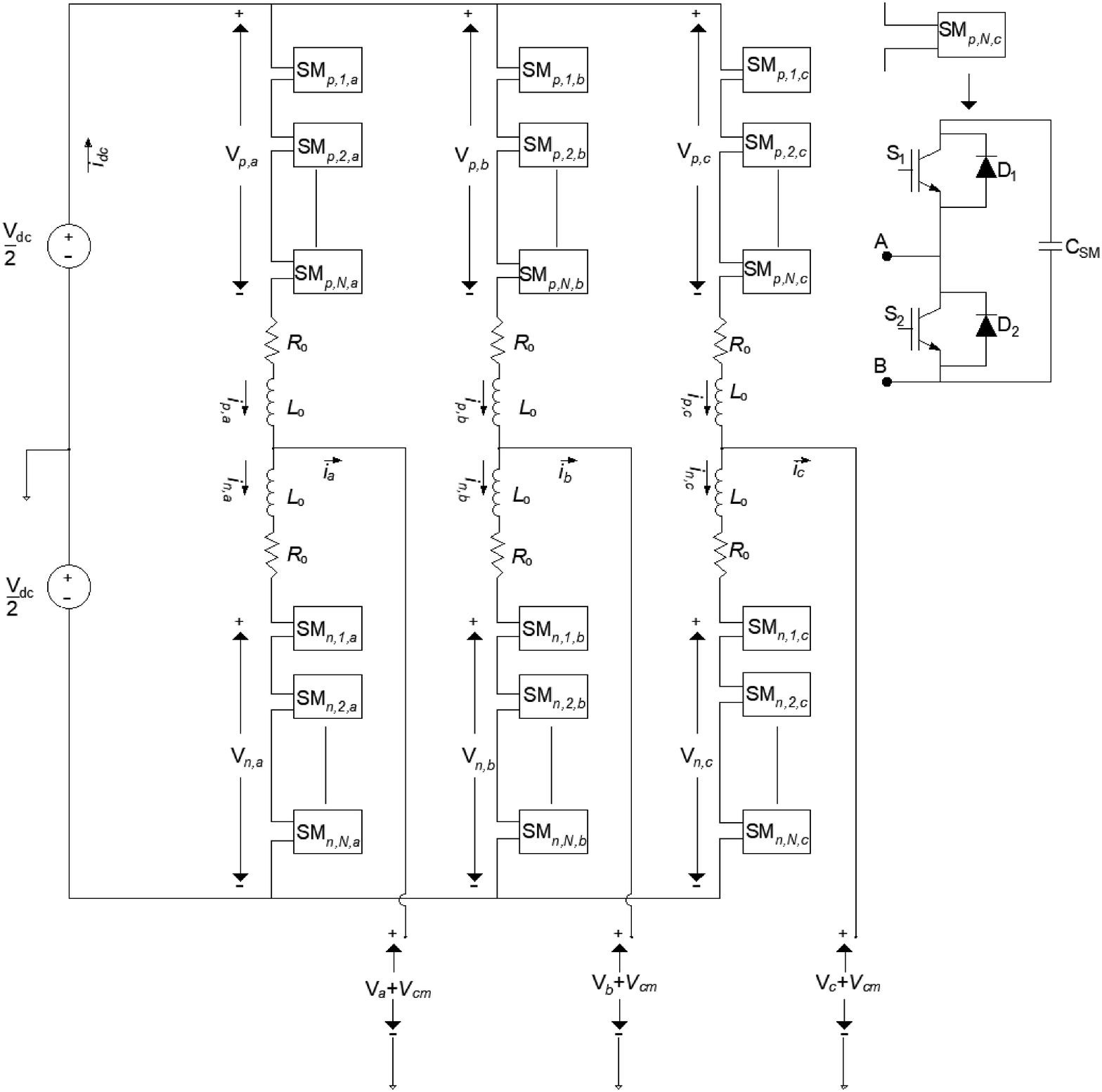}
    \caption{Detailed Schematic of MMC}
    \label{fig:MMC detail}
\end{figure}

Such advantages can be achieved through the implementation of Voltage Source Converter (VSC) HVdc networks \cite{Hertem2010}. HVdc networks have also been found to be beneficial in interlinking multiple ac asynchronous generation systems with the help of underground cables and over head transmission lines. The modular multi-level converter (MMC) has emerged as a popular choice for VSC-HVdc systems, due to certain salient features including (1) the absence of large dc link capacitors; (2) better scalability; (3) higher operational efficiency, etc. \cite{Debnath2015}. Conventional MMC design, as shown in Fig. \ref{fig:MMC detail}, uses half-bridge submodules (HBSM) rather than full-bridge modules (FBSM). FBSMs are fault tolerant, but have lower operational efficiency than HBSMs due to higher number semiconductor switches.

Given the remote locations of HVdc lines it is a challenge to detect faults early enough to prevent instability. Rapid isolation of faults is essential as it might cause indelible damaged to the converter stations and the network infrastructure \cite{Mitra2017}. Extensive research has been carried out for ac transmission systems for fault location, but such techniques are generally not applicable for HVdc systems \cite{Yang2019}.\par

Traditional ac systems utilize phasor and voltage angle information from Phasor Measurement Units (PMU) \cite{Ying-HongLin2004}, for fault location. Lack of phasor information and frequency data makes it difficult to use those methods. Multiple techniques have been used for identification of fault locations of dc lines. They can be broadly classified into two categories (1) single ended measurements and (2) double ended measurements \cite{Magnago1998}. Current approaches mostly discuss the use of time-domain based fault-location algorithms \cite{Nanayakkara2012}, \cite{Shang2001}. The high frequency fault transients contain information about the fault and its characteristics \cite{Mitra2019}. These traveling wave based methods have gained prominence due to the presence of time synchronized Global Positioning System (GPS) devices. These devices are expensive and their accuracy is dependent on the ability to capture the arrival of wave peaks \cite{He2014}. Accurate detection of the traveling waves are also dependent of the length of the line. Traveling waves are not affected by fault resistance, system parameters etc. \cite{Evrenosoglu2005}. Single ended measurements are cheaper but they tend to provide inaccurate results as the devices must have the capability to detect the reflected peak \cite{Xu2011}. The reflected surge waves are usually weak making it difficult to detect. The wave speed has an influence on the fault location accuracy. The surge propagation of the waves are dependent on the line parameters, controlling the accuracy of the results \cite{SHULTZ1986103}. Modern methods also involve the use of digital signal processing methods requiring high sampling frequency to achieve the desired accuracy \cite{Mitra2019}. Simultaneous time-frequency based methods like wavelet transform has been widely used for fault location. Discrete wavelet transform (DWT) and continuous wavelet transform (CWT) methods have been implemented. CWT tends to provide better resolution as compared to DWT. CWT involves a smooth shift of the mother wavelet over the time-domain, whereas in DWT, the mother wavelet is shifted using a dyadic pattern over time \cite{Mitra2019}, \cite{Vetterli1992}. Active fault location detection techniques using external injection of voltages using a power probe unit (PPU) has been suggested in \cite{Park2013} or as a pre-charged capacitor connected to a circuit breaker \cite{Liu2019}. The external oscillation circuit injects a signal whose under-damped oscillation and attenuation is used to locate the fault. The requirement of an external probe unit or a pre-charged capacitor has been suggested for low voltage dc (LVdc) networks \cite{Mohanty2016}. They are difficult to achieve for large HVdc networks, as discharging a pre-charged external capacitance into the network can cause over voltage problems and can damage the infrastructure.\par

Use of Artificial Intelligence (AI) to locate faults and also improve its accuracy has been previously suggested in the literature \cite{Novosel1996}. Measured voltage and current are utilized as inputs to the neural network. The corresponding features are utilized for fault location on transmission systems. Other methods using statistical data classifications like Support Vector Machine (SVM) have been studied in \cite{Dash2007} for fault location in transmission lines. The data driven methods for fault location requires a training dataset and the neural network has to be retrained for every new dataset, that requires significant time and effort, and is computationally burdensome.\par

In this paper, a passive method for fault location using the natural attenuation of the transmission line current following the isolation of the fault, is suggested. After fault isolation, the transmission line capacitance discharges into the fault through the line inductance and resistance. Under the absence of any active voltage source, the damped response of the transmission line current provides us with the rate of attenuation of the fault current. This information along with the damped natural frequency of the transmission line current calculated using Fast Fourier Transform (FFT) analysis is suggested in the paper for fault location. The attenuation constant of the damped transmission line current is calculated using the linear regression (LR) method \cite{Zhang2018}. The paper also investigates fault isolation using a hybrid dcCB and then the damped natural frequency of the transmission line current helps to provide the fault location. A double ended local measurement is utilized for better accuracy in fault location. The passive method of fault location is achieved without any signal injection or external circuits, thereby reducing costs and complexities associated with it.\par

The accuracy of the proposed method is verified under varying fault locations and fault resistances. Sensitivity to measurement noise and other parameters is performed in a three terminal MTdc network. The rest of the paper is organized as follows: Section \ref{Sec:Modeling} discusses the modeling of the MTdc network; Section \ref{Sec:Method} discusses the proposed fault location methodology; Section \ref{Sec:Simulation Results} discusses the performance of the fault location methodology through simulation studies performed in PSCAD/EMTdc; Section \ref{Discussion} discusses the contributions and finally Section \ref{Conclusion} concludes the paper with major findings.
\vspace{-1em}
\section{Modeling} \label{Sec:Modeling}
\subsection{The MMC}
The modeling of the three terminal MMC in this paper is based on the design suggested in \cite{Debnath2018}. The MMC model consists of 400 half bridge sub-modules (HBSM) per arm. Hybrid discretization and relaxation algorithms described in \cite{Debnath2018} are used to define the numerical stiffness in the differential algebraic equations. The control  of  the  MMCs  is based on the strategies explained in \cite{Debnath2016}, \cite{Debnath2017}. More details about the system parameters are provided in Table \ref{Table 1}.

\begin{figure}
    \centering
    \includegraphics[width=0.5\textwidth]{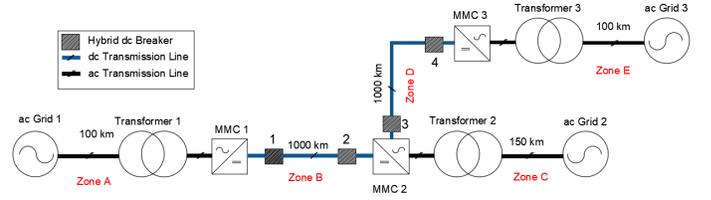}
    \caption{MTdc protection zones}
    \label{fig:zones}
\end{figure}

\begin{table}[]
\centering
\caption{System Parameters}\label{Table 1}

\begin{tabular}{|c|c|c|}
\hline
 & \textbf{Parameters} & \textbf{Value} \\ \hline
\multirow{5}{*}{\textbf{ac side}} & Voltage (L-L RMS) & 333 kV \\ \cline{2-3} 
 & Length of transmission line 1 \& 3 & 100 km \\ \cline{2-3} 
 & Length of transmission line 2 & 150 km \\ \cline{2-3} 
 & System Frequency & 60 Hz \\ \cline{2-3} 
 & Transmission line resistance & 0.03206 $\Omega$/km \\ \hline
\multirow{4}{*}{\textbf{dc side}} & Voltage (L-L) & 640 kV \\ \cline{2-3} 
 & Length of transmission line & 1000 km \\ \cline{2-3} 
 & Transmission line resistance & 0.03206 $\Omega$/km \\ \cline{2-3} 
 & MMC capacity & 1 GW  \\ \hline
\end{tabular}

\end{table}

\vspace{-1em}

\subsection{MTdc Network}
The model of a radial three-terminal MTdc symmetric monopole is shown in Fig. \ref{fig:zones}. The system is equipped with hybrid dcCB at the MMC terminals. The dc transmission lines are designed as frequency dependent models having 6 conductors with a vertical spacing of 5m and horizontal spacing of 10m between the conductors. 
\subsection{Hybrid dcCB}
\begin{figure}
    \centering
    \includegraphics[width=0.5\textwidth]{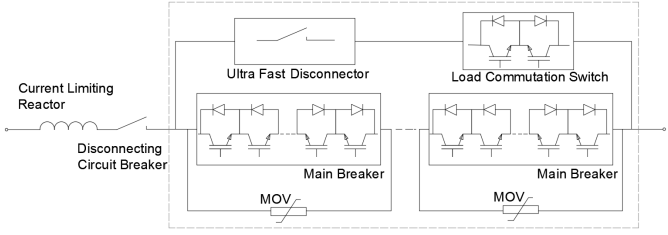}
    \caption{Schematic diagram of hybrid dc breaker}
    \label{fig:Hybrid_Breaker}
\end{figure}

The hybrid dc breaker model represents the breaker designed by ABB in \cite{Callavik2012}. The breaker design, as shown in Fig. \ref{fig:Hybrid_Breaker}, comprises of three major sections, (1) load commutation branch, (2) main breaker branch and (3) the energy absorption branch. Under normal operating conditions, the load commutation branch remains operational. As the fault detection command is generated, the load commutation switches turn-off, the current recedes through the ultra-fast disconnector into the main breaker branch. On current zero detection, the ultra-fast disconnect switch is opened. The fast mechanical switch protects the commutation switch from the entire dc line voltage during final interruption. The final interruption happens in the main breaker branch. The excess energy is absorbed by the Metal Oxide Varistor (MOV's). The maximum time required to dissipate the energy depends on the capacity of the MOV banks. The MOV's were designed and rated at $800$ kV. The introduction of two parallel branches reduces the on-state losses to 0.01\% of the transmitted power \cite{Mitra2017}, \cite{RayChaudhuri2014}. Faults occurring on the dc transmission line need to be interrupted very quickly. Current limiting inductors connected in series with the dc breakers act as protective devices for the switches, and limit the rate of change of fault current across them.

\section{Proposed Fault Location Methodology} \label{Sec:Method}
Fault detection is a challenge in MTdc systems and various research work have been reported in \cite{Mitra2019}, \cite{Sneath2016}, \cite{Li2017}. Once a fault is detected, the hybrid dcCBs' operate to isolate the faulted section of the network. The other non-faulted sections remain operational. The entire MTdc network has been divided into multiple protection zones as shown in Fig. \ref{fig:zones}. For a hybrid dcCB, faults occurring internally are within their zones of protection. Zone B and zone D are the internal zones of protection for dcCB 1 and 2, and dcCB 3 and 4 respectively. The formulation of the problem is carried out on a single phase long line shown in Fig. \ref{fig:eqvcable} (a). $R_k$, $L_k$ and $C_k$ are the series resistance, inductance and shunt capacitance, measured in $\Omega$, mH and $\mu$F per unit length respectively. It is assumed they are constant and are uniformly distributed along the line length. Line conductance is neglected. Once the fault current is interrupted, the stored energy in the transmission line capacitance discharges into the fault. The stored energy of the transmission line capacitance at a certain distance from the terminals is finite. In the absence of any external voltage source, the transmission line current $i_{line}$ discharges into the fault over time. Since the faulted section is isolated, MMC controls do not affect the current discharge. The rest of the isolated network upto the fault point can be considered a \emph{RLC} oscillating circuit, with the current resonating similarly as an \emph{LC} circuit and the presence of the resistance decays the oscillations over a period of time. By analyzing the discharging transmission line current $i_{line}(t)$, the fault location in the transmission line can be estimated.

\vspace{-1em}
\subsection{Faulted section formulation} \label{subsec:faulted}

\begin{figure}
    \begin{subfigure}{\linewidth}
    \includegraphics[width=\textwidth]{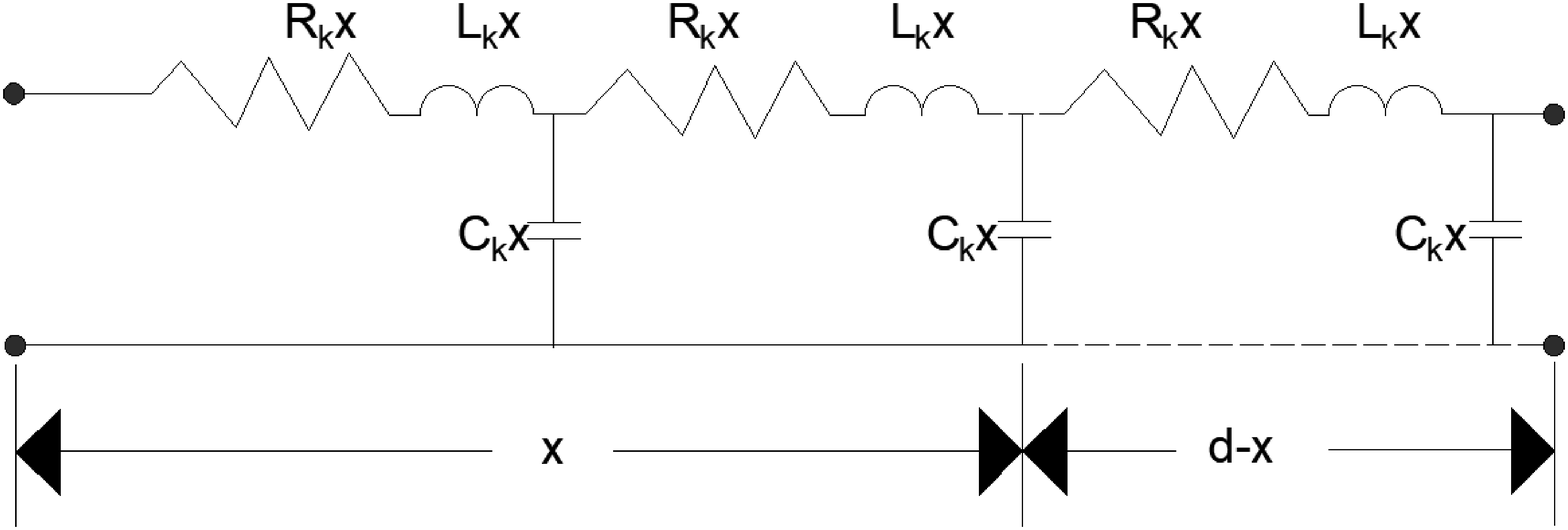}
    \caption{Unfaulted long line}
    \end{subfigure}
    \begin{subfigure}{\linewidth}
    \includegraphics[width=\textwidth]{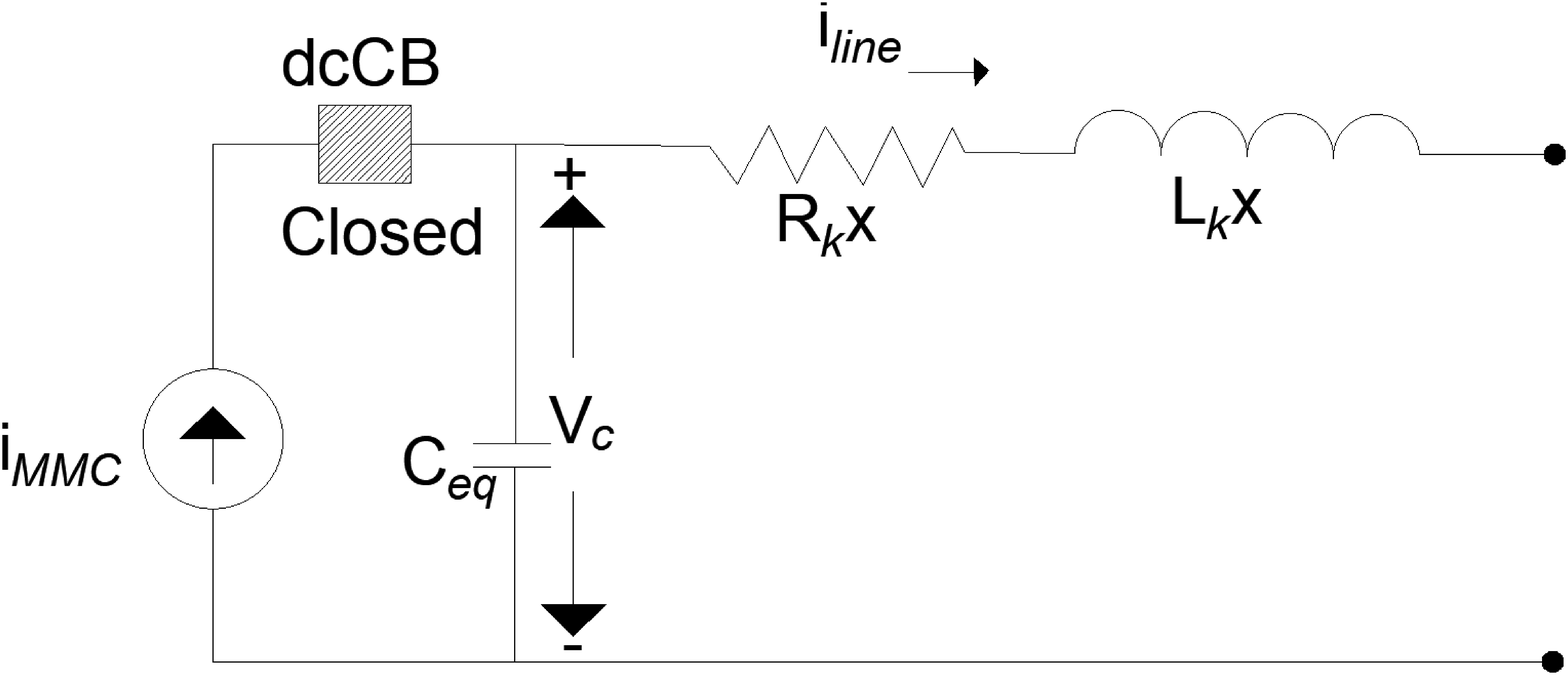}
    \caption{Normal operation with dcCB}
    \end{subfigure}
    \begin{subfigure}{\linewidth}
    \includegraphics[width=\textwidth]{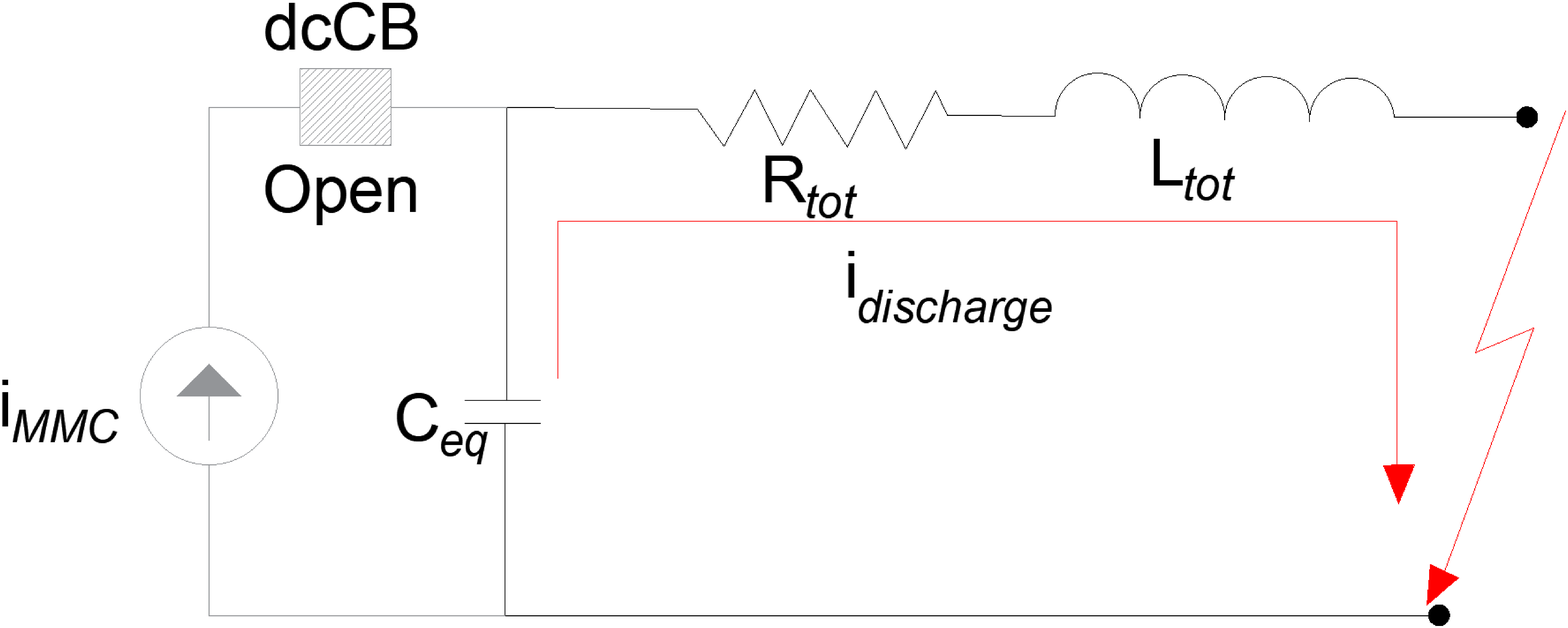}
    \caption{After fault detection and isolation}
    \end{subfigure}
    \caption{Equivalent transmission line representation}
    \label{fig:eqvcable}
\end{figure}

Once the faulted section of the transmission line is isolated, the remaining portion of the transmission line beyond the hybrid dcCB upto the fault point can be represented by an equivalent \emph{RLC} circuit as shown in Fig. \ref{fig:eqvcable}. Differential equations governing the state of the circuit can be calculated from Kirchhoff's Voltage Law (KVL), and the constitutive equations for the transmission line inductance, resistance and stored capacitance is given as (\ref{Eqn:KVL}),

\begin{equation}
    V_R+V_L+V_C=V(t) 
    \label{Eqn:KVL}
\end{equation}

where $V_R$, $V_L$ and $V_C$ are the voltages across the transmission line resistance, inductance and capacitance respectively. $V(t)$ is a time-varying voltage source. After the fault isolation in the absence of any time-varying voltage source $V(t) \rightarrow 0$. Substituting, $V_R=R_{tot} i_{line}(t)$, $V_L=L_{tot} \frac{\mathrm{d} i_{line}(t)}{\mathrm{d} t}$ and $V_C=\frac{1}{C_{eq}}\int_{0}^{t} i_{line}(t)dt$ in (\ref{Eqn:KVL}), we get (\ref{Eqn:KVL1}),
\begin{equation}
    R_{tot} i_{line}(t) + L_{tot}\frac{\mathrm{d} i_{line}(t)}{\mathrm{d} t} + \frac{1}{C_{eq}}\int_{0}^{t} i_{line}(t)dt = 0
    \label{Eqn:KVL1}
\end{equation}

Differentiating (\ref{Eqn:KVL1}) with time, we get a second order differential equation (\ref{Eqn:SecondOrder}),

\begin{equation}
\centering
    \frac{\mathrm{d}^2 i_{line}(t) }{\mathrm{d} t^2} +\frac{R_{tot}}{L_{tot}}\frac{\mathrm{d} i_{line}(t)}{\mathrm{d} t} + \frac{1}{L_{tot}C_{eq}}i_{line}(t)=0
    \label{Eqn:SecondOrder}
\end{equation}
Now $R_{tot}$ is the equivalent resistance upto the fault path including the transmission line resistance $R_{line}$ and the fault resistance $R_{fault}$. $L_{tot}$ is the net transmission line inductance and $C_{eq}$ is the equivalent capacitance from the point of consideration.\par
A more general solution to (\ref{Eqn:SecondOrder}) can be given as (\ref{Eqn:General}),
\begin{equation}
    \frac{\mathrm{d}^2 i_{line}(t) }{\mathrm{d} t^2} +2\alpha\frac{\mathrm{d} i_{line}(t)}{\mathrm{d} t} + \omega_0^2i_{line}(t)=0
    \label{Eqn:General}
\end{equation}

where $\omega_0$ is the undamped resonance frequency of oscillation, the rate at which the oscillation decays is determined by the attenuation $\alpha$ and they are represented as (\ref{Eqn:alpha}),
\begin{equation} \label{Eqn:alpha}
    \alpha=\frac{R_{tot}}{2L_{tot}};\: \omega_0=\frac{1}{\sqrt {L_{tot}C_{eq}}}
\end{equation}

\begin{figure}
    \centering
    \includegraphics[width=0.5\textwidth]{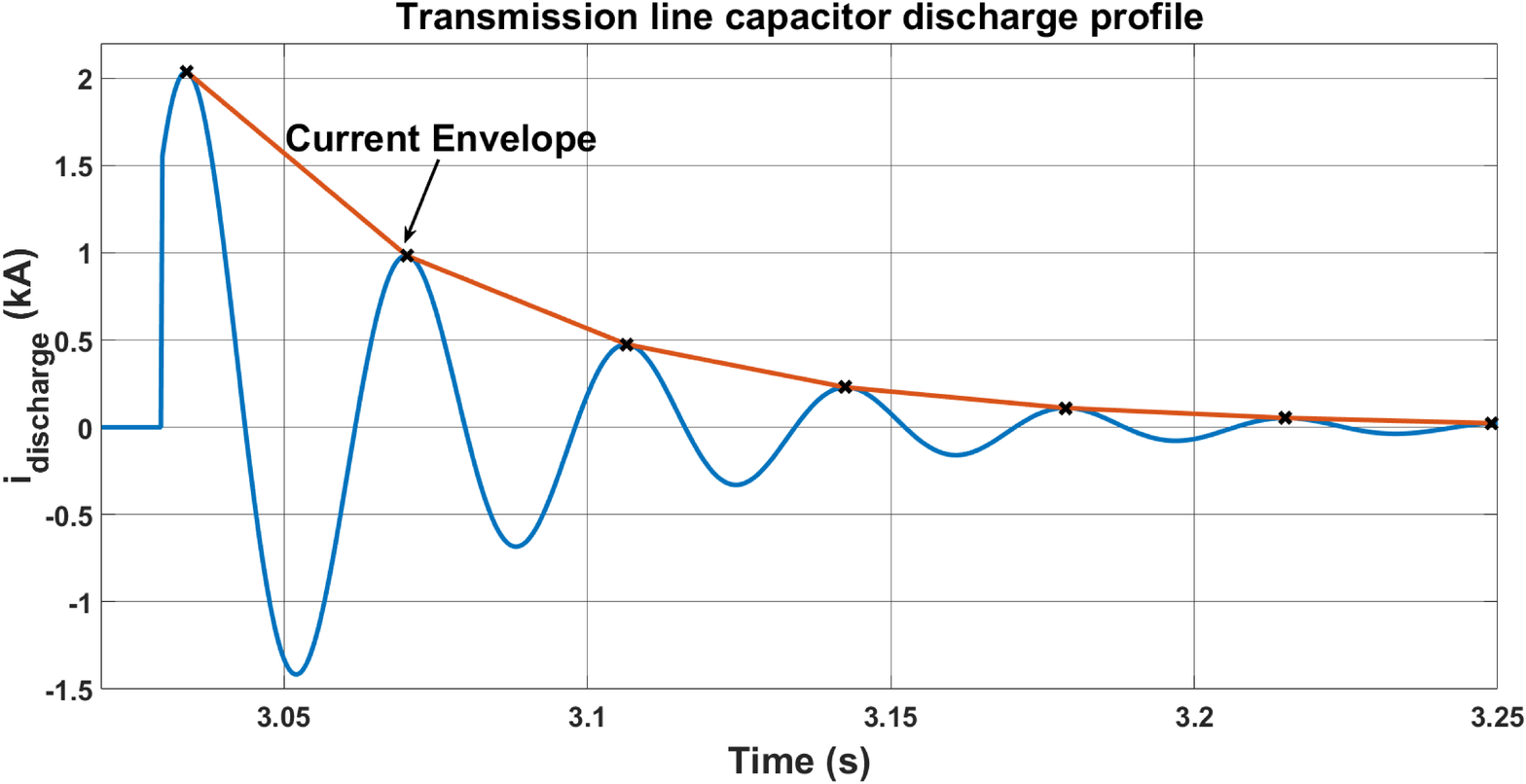}
    \caption{Capacitor discharge current profile in a transmission line after fault isolation}
    \label{fig:CapDischarge}
\end{figure}

As stated previously, in the absence of a driving voltage source other than the discharging transmission line capacitance, the solution for the transmission line current $i_{line}$ can be given as an under-damped response for a \emph{RLC} circuit. The general solution for the under-damped response is given as (\ref{Eqn:GenSol}),
\begin{equation}
    i_{line}(t) = D_1 e^{-\alpha t} cos(\omega_d t) + D_2 e^{-\alpha t} sin(\omega_d t)
    \label{Eqn:GenSol}
\end{equation}

where $\omega_d$ is the damped natural frequency of the capacitor and $\alpha$ is the rate at of attenuation of the stored energy of the capacitor. An example of the discharge profile of capacitor current in the transmission line after the dcCB has operated is shown in Fig. \ref{fig:CapDischarge}.

\vspace{-1em}
\subsection{Attenuation Constant}

The capacitor discharge upto the faulted point is an under-damped response. The attenuation constant can be calculated from the discharge current by considering the envelope of the oscillating signal. The measured under-damped oscillating current is sampled at regular intervals to obtain peaks of the signal. The envelope of the under-damped oscillating current can be represented as (\ref{Eqn:Envelope});

\begin{equation}
    i_{discharge}(t_n)=i_{peak}(t_n)e^{-\alpha t_n}
    \label{Eqn:Envelope}
\end{equation}

Taking natural logarithm, the above equation can also be represented as (\ref{Eqn:LR}),
\begin{equation}
    ln (i_{discharge}(t_n)) = ln (i_{peak}(t_n))- \alpha t_n
    \label{Eqn:LR}
\end{equation}
Thus, (\ref{Eqn:LR}) takes the form of a straight line (\ref{Eqn:Line}),
\begin{equation}
    \begin{matrix}
y=mx+c\\
y = ln(i_{discharge}(t)); m = -\alpha; c = ln(i_{peak}(t))
\end{matrix}
\label{Eqn:Line}
\end{equation}

Since (\ref{Eqn:LR}) can be represented as (\ref{Eqn:Line}), a linear regression (LR) approach can be considered to compute the slope of the line i.e, $\alpha$. From the data obtained by sampling $i_{discharge}$ at regular intervals the unknown model parameter can be estimated. For a given set of observations the model takes the form (\ref{eqn:LRform}),

\begin{equation}
    \begin{matrix}
    y_1= c_0 + m_1x_1 \\
    y_2 = c_0+m_2x_2 \\
    y_3 = c_0+ m_3x_3 \\
    .... \\
    y_i = c_0+ m_ix_i \\
   where, i=1,2,...,n
    \end{matrix}
    \label{eqn:LRform}
\end{equation}

The equivalent matrix form of (\ref{eqn:LRform}), can be written as (\ref{eqn:LRmat}),

\begin{equation}
   \begin{matrix}
   y=Ax^T \\
   \\
    y = \begin{bmatrix}
    y_1\\ 
    y_2\\ 
    y_3\\
    ..\\
    y_i 

\end{bmatrix} ;
    x^T = \begin{bmatrix}
    1 & x_1\\ 
    1 & x_2\\ 
    1 & x_3\\
    .. & ..\\
    1 & x_i 

\end{bmatrix};
    A = \begin{bmatrix}
   c\\ 
   m

\end{bmatrix}
\end{matrix}
\label{eqn:LRmat}
\end{equation}

Here $y$ is is the set of observed variables at different time-steps, $x$ is the set of exogenous or input variables. Solving (\ref{eqn:LRmat}), we calculate the attenuation $\alpha$ from the entries of matrix $A$.
\vspace{-1em}
\subsection{Fast Fourier Transform (FFT)} \label{subsec:FFT}

One can transform a given sequence in time into its respective frequency components using Discrete Fourier Transform (DFT) \cite{He2016}. FFT is useful to perform the DFT of a sequence. FFT performs the computation of the DFT matrix as a product of sparse factors. The DFT for such a sequence can be given as (\ref{Eqn:DFT}),

\begin{equation}
    X[k] = \sum_{n= 0}^{N-1}x[n]e^{-j2\pi kn /N}
    \label{Eqn:DFT}
\end{equation}

where \emph{N} is the length of the signal. Since, the sampling frequency of the signal is varied between 10 kHz to 100 kHz, the maximum represented frequencies are half of the sampling frequency. We try to capture all the representative frequencies in that range. The peak amplitude of the damped natural frequency is $\omega_d$. To calculate the damped natural frequency of the capacitor discharge $\omega_d$, we perform FFT analysis to determine the dominant frequency of the under-damped oscillating transmission line current. As stated earlier, the transmission line current $i_{discharge}(t)$ decays at a frequency of $\omega_d$ as shown in (\ref{Eqn:GenSol}). The undamped response of the decaying oscillation can be calculated as (\ref{eqn:DampedFreq}),

\begin{equation}
    \omega_d = \sqrt{\omega_0^2 - \alpha^2}
    \label{eqn:DampedFreq}
\end{equation}

The damping factor $\zeta$ is given as a ratio of $\omega_d$ and $\alpha$ as (\ref{Eqn:Zeta}),
\begin{equation}
    \omega_d = \omega_0\sqrt{1-\zeta^2}
    \label{Eqn:Zeta}
\end{equation}

where, 
\begin{equation*}
 \zeta = \frac{\alpha}{\omega_0}=\frac{R_{tot}}{2}\sqrt{\frac{C_{eq}}{L_{tot}}}   
\end{equation*}

\subsection{Fault Location Calculation}

From the PSCAD transmission line modelling parameters, the per unit (p.u) length line inductance ($L_k$), resistance $(R_k)$ and the per unit capacitance ($C_{k}$) from the point of measurement can be calculated. The undamped resonance frequency of oscillation $\omega_0$ of the capacitor discharge current through the transmission line can be calculated as (\ref{Eqn:NaturalFreq});

\begin{equation}
    \omega_0^2 = \omega_d^2 + \alpha^2
    \label{Eqn:NaturalFreq}
\end{equation}

Thus, fault location \emph{$d_{cal}$} can be calculated as (\ref{Eqn:Distance}),

\begin{equation}
    d_{cal} = \frac{1}{(\omega_d^2 + \alpha^2)L_k C_k}
    \label{Eqn:Distance}
\end{equation}

where, $L_{tot} = L_k d_{cal}$. To verify the robustness of the algorithm, faults are created at various length of the transmission line and the fault resistance is also varied between $0.01\Omega$ and $200\Omega$. The error $\%$ between the actual fault location and the measured location is given by $\epsilon$ \cite{Gopal2000}, 
$d_{act}$ is the actual location of the fault in the transmission line and $d_{cal}$ is the calculated fault location using (\ref{Eqn:Distance}).
\vspace{-1 em}
\subsection{Proposed Algorithm}

The following section discusses, in brief, the proposed algorithm to detect the fault location, Algorithm \ref{Algo1} shows the steps in the process.

Local measurements of current and voltage are sampled at each location in real-time. At the onset of the fault, the traveling waves are detected that help to isolate the faulted section of the transmission line. The fault detection algorithm is robust to various changes of operating conditions and measurement noise. After the breakers operate, the current recorder devices starts to monitor the transmission line discharge current. The recorded current $i_{discharge}$ is sampled at every 2ms to determine the peaks of the current envelope $I_{0peak}$. A FFT analysis is also performed on $i_{discharge}$ to determine the damped natural frequency. Finally, the fault is located using (\ref{Eqn:Distance}).

\begin{algorithm}
\caption{Fault location using transmission line current discharge}
\label{Algo1}
\SetAlgoLined
\DontPrintSemicolon

Sampling Frequency ($f_s$)= 10kHz, i.e., $\triangle$t= 0.1ms;

Fault occurs at T = $t_s$;\

At $T$= $t_{brk}$ (Hybrid dcCB operates):\;
Hybrid dcCB operate at their respective zones to isolate the fault;\;
\eIf {$Breaker_{Status}=0$ (Open)}{
Enable fault location algorithm;\;
Measure and store $i_{line}$ at $\triangle$t as $i_{discharge}$;\;
Sample $i_{discharge}$ at 2ms; the peak value of $i_{discharge}$ is stored as $i_{peak}$ from (\ref{Eqn:Envelope});\;
Calculate $\alpha$ using LR method from (\ref{eqn:LRmat});\;
FFT of $i_{discharge}$ to extract $\omega_d$ using (\ref{Eqn:DFT});\;
Calculate the location of fault using (\ref{Eqn:Distance});\;
}{
\textbf{Continue} Normal operation;}

\end{algorithm}
 \vspace{-1.1em}
\section{Simulation Results} \label{Sec:Simulation Results}

\begin{figure*}
    \centering
    \includegraphics[width=1\textwidth]{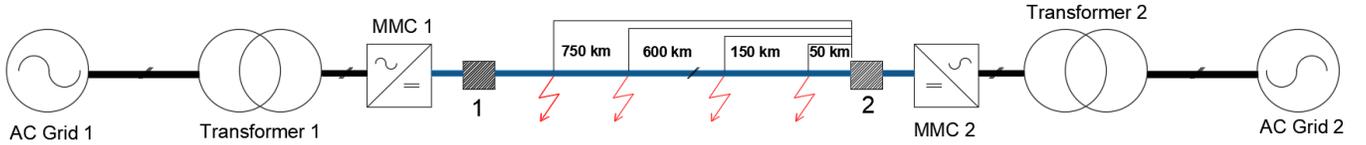}
    \caption{Fault locations}
    \label{fig:locations}
\end{figure*}

To verify the accuracy of the proposed algorithm the radial MTdc network shown in 
Fig. \ref{fig:zones} was designed in PSCAD/EMTdc. Varying simulation conditions like fault distances and fault resistances were performed on the 1000km long dc section of the transmission line. As discussed previously, the proposed method does not require any external injection of current or voltage pulses into the network as proposed in \cite{Park2013,Liu2019,Mohanty2016}. The fault detection algorithm detects the faults in the zone of internal protection as described in Section \ref{Sec:Method}. The fault locations obtained by the proposed method was compared to the actual fault location and an error metric is calculated. A schematic for various fault locations is shown in Fig. \ref{fig:locations}.
\begin{figure}
    \centering
    \includegraphics[width=0.5\textwidth]{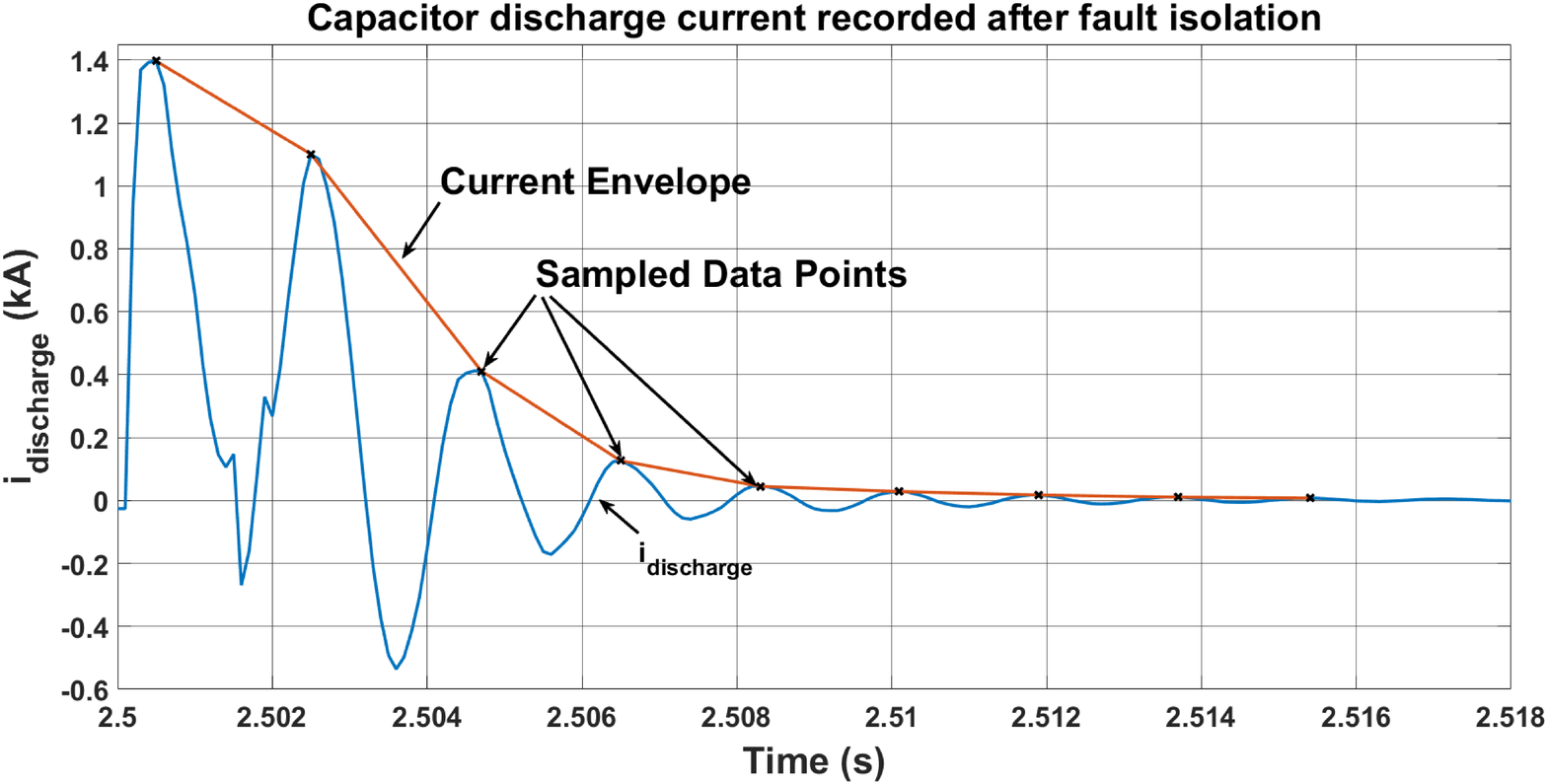}
    \caption{Capacitor discharge current with current envelope for 0.01 $\Omega$ fault at 50 km from MMC 1}
    \label{fig:discharge}
\end{figure}

\begin{figure}
    \centering
    \includegraphics[width=0.5\textwidth]{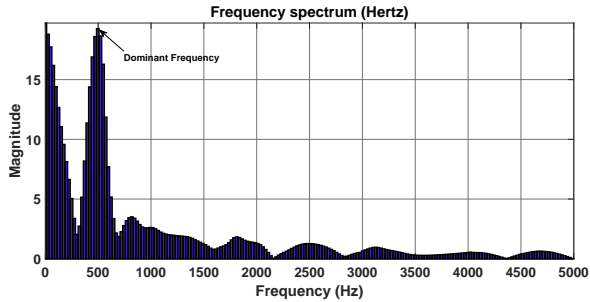}
    \caption{FFT analysis of $I_{discharge}$ for 0.01 $\Omega$ fault at 50 km from MMC 1}
    \label{fig:fft}
\end{figure}

\begin{figure}[ht!]
    \centering
    \includegraphics[width=0.5\textwidth]{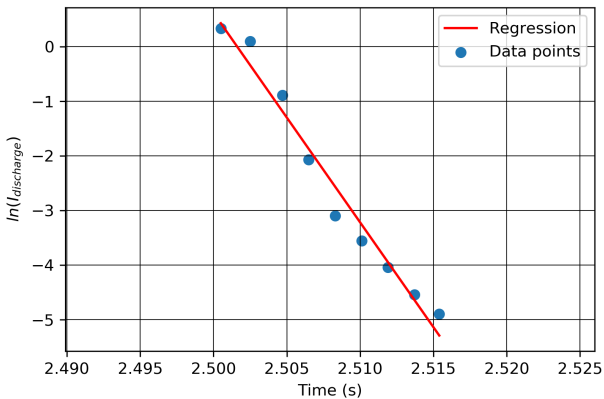}
    \caption{Regression model for 0.01 $\Omega$ fault at 50 km from MMC 1}
    \label{fig:predict}
\end{figure}

A pole-to-ground fault was simulated in zone B, hybrid dcCBs' 1 and 2 operated to isolate the fault. The actual location of the fault was 50km from the recorder located at a distance from MMC 2. Fig. \ref{fig:discharge} shows the sampled discharge current at the recorder location. A FFT analysis for the sampled current is shown in Fig. \ref{fig:fft}. Using the sampled data points from the current envelope we determine $\alpha$ from (\ref{eqn:LRmat}). Coefficient of determination that is used as a statistical measure for the performance of the regression model for the data, is found to be $R^2 = 0.9722$, indicating a high degree of linear relationship for the straight line regression model as explained in (\ref{Eqn:Line}). Using the equations to calculate the  actual fault distance was found with \%$\epsilon$ of 0.007. The predicted linear regression model plotted against the measured data is shown in Fig. \ref{fig:predict}.

Similar events of fault were performed across the transmission line as shown in Fig. \ref{fig:locations}.
Table \ref{Table: Fault Dist_Nonoise} shows the various fault distance estimations, error and the coefficient of determination for the calculated regression model for $\alpha$.

\begin{table}
\centering
\caption{Fault distance estimation, without measurement noise}
\label{Table: Fault Dist_Nonoise}
\begin{tabular}{|c|c|c|c|}
\hline
\textbf{\textit{$d_{act} (km)$}} & \textit{\textbf{$R_{fault}$ ($\Omega$)}} & \textit{\textbf{$\epsilon$ (\%)}} & \textit{\textbf{$R^2$}} \\ \hline
50 & 0.01 & 0.0070 & 0.9722 \\ \hline
150 & 0.01 & 0.0120 & 0.9641 \\ \hline
600 & 0.01 & 0.0141 & 0.8399 \\ \hline
750 & 0.01 & 0.0166 & 0.9649 \\ \hline
50 & 2 & 0.0020 & 0.9796 \\ \hline
150 & 2 & 0.0113 & 0.9795 \\ \hline
600 & 5 & 0.0120 & 0.8345 \\ \hline
750 & 5 & 0.0148 & 0.9773 \\ \hline
50 & 10 & 0.0024 & 0.9821 \\ \hline
150 & 10 & 0.0188 & 0.9667 \\ \hline
600 & 10 & 0.0133 & 0.8145 \\ \hline
750 & 10 & 0.0150 & 0.9768 \\ \hline
150 & 50 & 0.0114 & 0.9745 \\ \hline
200 & 50 & 0.0110 & 0.9858 \\ \hline
150 & 100 & 0.0200 & 0.9783 \\ \hline
200 & 100 & 0.0210 & 0.9446 \\ \hline
150 & 200 & 0.0232 & 0.9758 \\ \hline
600 & 200 & 0.0350 & 0.9438 \\ \hline
\end{tabular}
\end{table}

Variations in fault resistance between $R_{fault} = 0.01\Omega$ to $200\Omega$, was performed. The fault locations were also varied and the algorithm was tested for faults in zone B and D.\par

\begin{figure}
    \centering
    \includegraphics[width=0.5\textwidth]{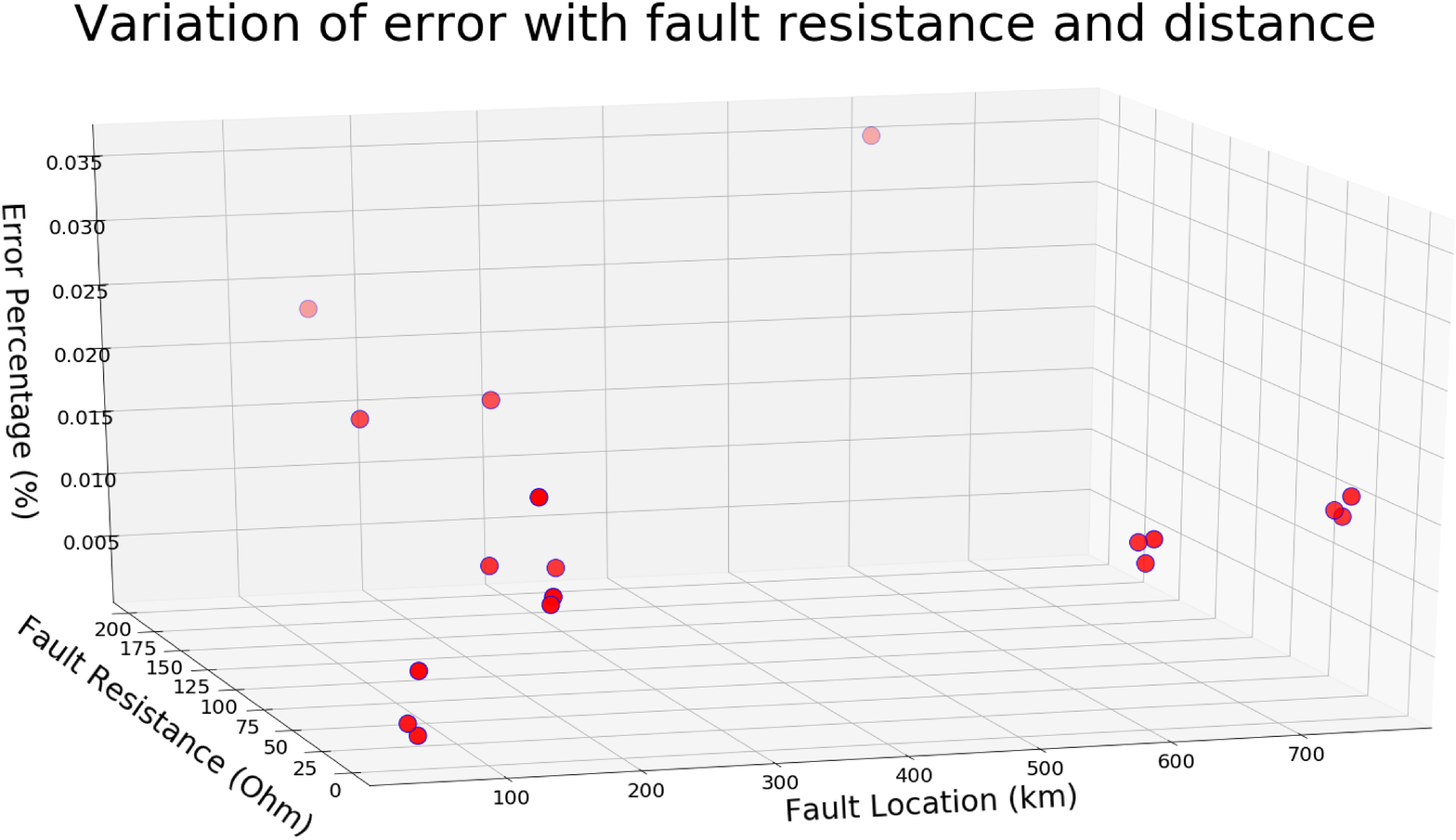}
    \caption{Variation of error with $R_{fault}$ and location}
    \label{fig:3Dplot}
\end{figure}

Variation in error with changes in $R_{fault}$ and $d_{act}$ are shown in Fig. \ref{fig:3Dplot}. From Table \ref{Table: Fault Dist_Nonoise} it can be seen that the \% error increases ever so slightly with the increment of $R_{fault}$, as the capacitor discharge attenuates at a faster due to the presence of a larger fault resistance. The rate of attenuation is affected, that causes the \% error to vary.

\vspace{-1em}
\subsection{Measurement Noise}
\begin{figure}
    \centering
    \includegraphics[width=0.5\textwidth]{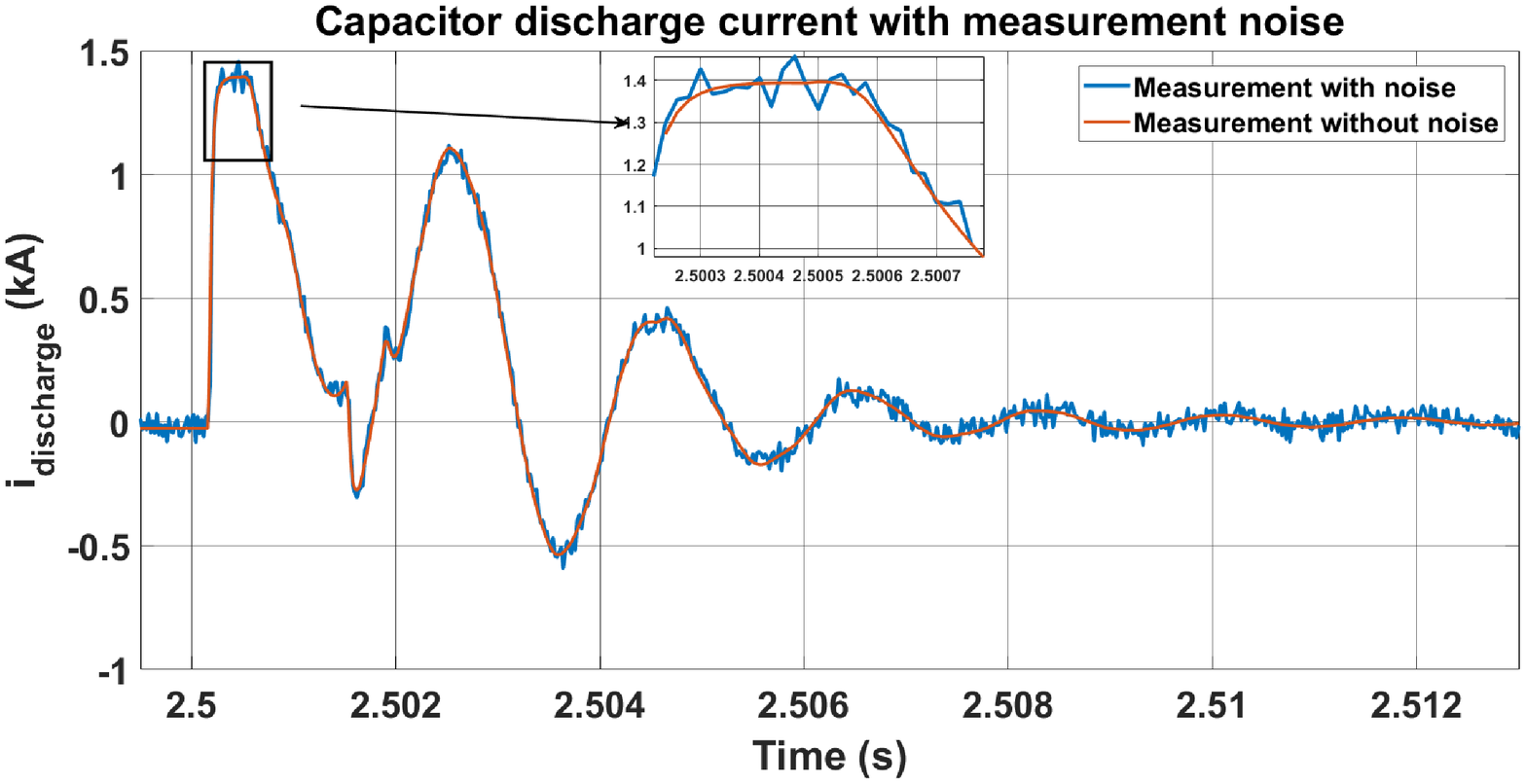}
    \caption{Capacitor discharge current with measurement noise}
    \label{fig:noise}
\end{figure}
To verify the influence of the real field measurement noise on the fault location estimation, a Gaussian noise spectrum of SNR 30dB, with a standard deviation of 2\%, was added to the measured data as suggested in \cite{Liu2011}. Fig. \ref{fig:noise} shows a current discharge profile with added measurement noise. The fault location estimation is summarized in Table \ref{Table: Fault Dist_noise}. The calculation of the attenuation constant $\alpha$ under the influence of measurement noise is not affected. 
\vspace{-1em}

\begin{table}[h]
\centering
\caption{Fault distance estimation, with 30dB measurement noise}
\label{Table: Fault Dist_noise}
\begin{tabular}{|c|c|c|c|}
\hline
\textbf{\textit{$d_{act} (km)$}} & \textit{\textbf{$R_{fault}$ ($\Omega$)}} & \textit{\textbf{$\epsilon$ (\%)}} & \textit{\textbf{$R^2$}} \\ \hline
50 & 0.01 & 0.010 & 0.9712 \\ \hline
150 & 0.01 & 0.0140 & 0.9641 \\ \hline
600 & 0.01 & 0.0131 & 0.8399 \\ \hline
750 & 0.01 & 0.0156 & 0.9649 \\ \hline
50 & 2 & 0.0033 & 0.9796 \\ \hline
150 & 2 & 0.0144 & 0.9715 \\ \hline
600 & 5 & 0.0135 & 0.8040 \\ \hline
750 & 5 & 0.0205 & 0.9743 \\ \hline
50 & 10 & 0.0030 & 0.9832 \\ \hline
150 & 10 & 0.0192 & 0.9667 \\ \hline
600 & 10 & 0.0153 & 0.8144 \\ \hline
750 & 10 & 0.0156 & 0.9745 \\ \hline
200 & 50 & 0.0127 & 0.9543 \\ \hline
150 & 100 & 0.0236 & 0.9687 \\ \hline
200 & 100 & 0.0261 & 0.9332 \\ \hline
150 & 200 & 0.0232 & 0.9748 \\ \hline
600 & 200 & 0.0387 & 0.9422 \\ \hline
\end{tabular}
\end{table}

\subsection{Sampling Frequency}
For the method of locating the fault on the transmission line the data was sampled at $f_s$ = 10kHz. To verify the effects of changes in sampling frequency over the observed peaks of discharge current is studied in this section. A similar application was performed with sampling frequencies of $f_s$= 50kHz and $f_s$= 100kHz, the results of the studies are summarized in \ref{Table: Fault Dist_20} and Table \ref{Table: Fault Dist_10} respectively.

\begin{table}[h]
\centering
\caption{Fault distance estimation, sampling at 50kHz with 30dB measurement noise}
\label{Table: Fault Dist_20}
\begin{tabular}{|c|c|c|c|}
\hline
\textbf{\textit{$d_{act} (km)$}} & \textit{\textbf{$R_{fault}$ ($\Omega$)}} & \textit{\textbf{$\epsilon$ (\%)}} & \textit{\textbf{$R^2$}} \\ \hline
50 & 0.01 & 0.006 & 0.9632 \\ \hline
150 & 0.01 & 0.0114 & 0.9655 \\ \hline
600 & 0.01 & 0.0136 & 0.8543 \\ \hline
750 & 0.01 & 0.0168 & 0.9865 \\ \hline
50 & 2 & 0.0031 & 0.9875 \\ \hline
150 & 2 & 0.0159 & 0.9833 \\ \hline
600 & 5 & 0.0143 & 0.8243 \\ \hline
750 & 5 & 0.0165 & 0.9767 \\ \hline
50 & 10 & 0.0032 & 0.9826 \\ \hline
750 & 10 & 0.0192 & 0.9345 \\ \hline
150 & 100 & 0.0236 & 0.9671 \\ \hline
200 & 100 & 0.0235 & 0.9322 \\ \hline
150 & 200 & 0.0282 & 0.9750 \\ \hline
600 & 200 & 0.0377 & 0.9520 \\ \hline
\end{tabular}
\end{table}

\begin{table}[h]
\centering
\caption{Fault distance estimation, sampling at 100kHz with 30dB measurement noise}
\label{Table: Fault Dist_10}
\begin{tabular}{|c|c|c|c|}
\hline
\textbf{\textit{$d_{act} (km)$}} & \textit{\textbf{$R_{fault}$ ($\Omega$)}} & \textit{\textbf{$\epsilon$ (\%)}} & \textit{\textbf{$R^2$}} \\ \hline
50 & 0.01 & 0.0024 & 0.9565 \\ \hline
150 & 0.01 & 0.0167 & 0.9555 \\ \hline
600 & 0.01 & 0.0137 & 0.8297 \\ \hline
750 & 0.01 & 0.0154 & 0.9516 \\ \hline
50 & 2 & 0.0031 & 0.9755 \\ \hline
150 & 2 & 0.0126 & 0.9634 \\ \hline
600 & 5 & 0.0153 & 0.8309 \\ \hline
750 & 5 & 0.0249 & 0.9676 \\ \hline
50 & 10 & 0.0043 & 0.9832 \\ \hline
750 & 10 & 0.0165 & 0.9438 \\ \hline
150 & 100 & 0.0239 & 0.9667 \\ \hline
200 & 100 & 0.0242 & 0.9552 \\ \hline
150 & 200 & 0.0253 & 0.9640 \\ \hline
600 & 200 & 0.0387 & 0.9557 \\ \hline
\end{tabular}
\end{table}
\vspace{-1.1em}
\section{Discussion} \label{Discussion}

The proposed method for identification of fault location has been achieved using a linear regression method to calculate the attenuation constant for the under-damped current oscillation. After successful fault detection and isolation, the proposed methodology is utilized for fault location on the affected transmission line. Robustness of the proposed method against varying fault resistance and location has been verified. Measurement noise was added to the simulated data to mimic real field measurements. The error, even under the influence of measurement noise was within 1\%. The fault location accuracy did not differ from the measurements without the influence of noise.

Methods involving current or voltage discharge into the faulted circuit through a pre-charged capacitor requires external devices to achieve the objective. For large MTdc networks, this is not possible. Besides, the voltage discharge into the network through the capacitor can cause over-current fluctuations that can further damage the overhead transmission lines.

To verify the effects of the sampling frequency on the measured data, a sampling frequency of 50kHz and 100kHz was used. The addition of measurement noise on the data did not result in a significant difference in the estimated fault location.

A comparison with current existing methods exhibited in Table \ref{Table:Comaprison}, shows the overall superiority of the method.

\begin{table}[]
\caption{Comparison of proposed algorithm with existing methods}
\label{Table:Comaprison}
\begin{tabular}{|l|l|l|}
\hline
\multicolumn{1}{|c|}{\textbf{Attribute}} & \multicolumn{1}{c|}{\textbf{Proposed method}} & 
\multicolumn{1}{c|}{\begin{tabular}[c]{@{}c@{}}\textbf{Other existing}\\\textbf{methods}\end{tabular}}\\
\hline
Communication & \begin{tabular}[c]{@{}l@{}}Communication channels \\ are not required\end{tabular} & \begin{tabular}[c]{@{}l@{}}Communication channels\\ are a requirement\end{tabular} \\ \hline
Cost & \begin{tabular}[c]{@{}l@{}}Separate devices are not \\ required to process\\information\end{tabular} & \begin{tabular}[c]{@{}l@{}}Specially designed and GPS\\ synchronized devices are\\ required\end{tabular} \\ \hline
Complexity & \begin{tabular}[c]{@{}l@{}}Network alterations and \\ any external devices are \\ not required\end{tabular} & \begin{tabular}[c]{@{}l@{}}Most methods require\\ external devices to gather\\ data to estimate the fault\\ location\end{tabular} \\ \hline
Reliability & \begin{tabular}[c]{@{}l@{}}Current measurements are\\ recorded by the sensors that \\ are robust to noise and other \\ variations\end{tabular} & \begin{tabular}[c]{@{}l@{}}Reliability depends on the\\ accuracy  of capturing\\ the weak reflected peak\\ of the fault\end{tabular} \\ \hline
Accuracy & \begin{tabular}[c]{@{}l@{}}Lower for methods that require\\ more sophisticated \\ measurement devices\end{tabular} & \begin{tabular}[c]{@{}l@{}}High, since highly sensitive\\ devices are used to record\\ measurements\end{tabular} \\ \hline
\end{tabular}
\end{table}

\vspace{-1em}

\section{Conclusion} \label{Conclusion}

To accurately locate faults in a MTdc network, a method using the natural discharge of the transmission line current has been proposed. After a fault is detected in a particular zone the hybrid dcCBs' in the faulted section operate to isolate the faulted transmission line. Once the faulted transmission line is isolated, the rest of the network forms an under-damped oscillating \emph{RLC} circuit in the absence of a driving voltage source up to the fault point. A relationship between the damped natural frequency $\omega_d$ and the rate of attenuation $\alpha$ was established. The method of linear regression was utilized for calculation of the attenuation constant. The robustness of the proposed scheme was verified against various fault resistances and different fault locations. Measurement noise did not have much of an influence on the proposed method. The addition of noise causes variations in the peak measurements of the transmission line capacitor current discharge, but since the attenuation constant is calculated through fitting the observed data points through a linear regression model, $\alpha$ calculations are not affected. The fault location method proved quite robust for high resistance faults of upto 200$\Omega$. The maximum error recorded was 0.0387\%, for a high fault impedance and added 30dB measurement noise. The effect of variations in the sampling frequency was also investigated to ensure the correctness of the proposed method. This paper thus provided a successful passive fault location method without the injection of any external current or voltage signals.


\ifCLASSOPTIONcaptionsoff
  \newpage
\fi

\vspace{-1em}
\bibliographystyle{unsrt}
\bibliography{Fault_Loc.bib}
\balance

\end{document}